\documentstyle[11pt,epsf]{article}

\newcommand{\T}{\textstyle}

\newcommand{\diag}{\;\mbox{\rm\bf diag}}

\def\3{\ss{}}

\def\Heff{H_{\mbox{\tiny eff}}}
\def\d{{\rm d}}

\def\eps{\varepsilon}

\def\be{\begin{equation}}
\def\bea{\begin{eqnarray}}
\def\ee{\end{equation}}
\def\eea{\end{eqnarray}}

\def\bean{\begin{eqnarray*}}
\def\eean{\end{eqnarray*}}
\def\s2{\Sigma^2(l)}
\def\sg2{\Sigma^2_g(l)}
\def\Ct2{|C(t)|^2}
\def\Cgt2{|C_g(t)|^2}

\begin{document}
\bibliographystyle{unsrt}
\vspace{1.5cm}
\begin{center}
{\LARGE\bf
Correlations between resonances in a
 statistical scattering model}\\
\end{center}

\vspace{.9cm}

\begin{center}
{\large\sc T. Gorin$^{1,2}$, F.--M. Dittes$^4$, M. M\"uller$^3$, I. Rotter$^{1,2}$, 
T. H. Seligman$^3$} \\

\vspace{1.5cm}

{
$^1$Technische Universit\"at Dresden, Institut f\"ur Theoretische Physik,
D-01062 Dresden\\
$^2$ Forschungszentrum Rossendorf, Institut f\"ur Kern- und Hadronenphysik,
D-01314 Dresden  \\
$^3$ Centro Internacional de Ciencias, Cuernavaca, Mexico \\
$^4$ Max--Planck--Institut f\"ur Physik komplexer Systeme, D-01187 Dresden
} \end{center}

\vspace{2.5cm}

\begin{abstract}
The distortion of the regular motion in a quantum system by its coupling to the continuum of decay
channels is investigated. The regular motion is described by means of a Poissonian ensemble.
We focus on the case of only few channels $K<10$. The coupling to the continuum induces
two main effects, due to which the distorted system differs from a chaotic system 
(described by a Gaussian ensemble): 
1. The width distribution for large coupling becomes broader than the 
corresponding $\chi^2_K$ distribution in the GOE case. 
2. Due to the coupling to the continuum, correlations are induced not only between the positions
of the resonances but also between positions and widths. These correlations remain even in the 
strong coupling limit. In order to explain these results, an 
asymptotic expression for the width distribution is derived for the one channel case. It
relates the width of a trapped resonance state to the distance between its two neighboring levels.
\end{abstract}

\newpage

\section{Introduction}

Resonances are states of an open quantum mechanical system, in which
the internal dynamics of the underlying closed system is disturbed by the
coupling to the decay channels.
For small coupling strength the widths of the resonances increase with
growing coupling while their positions remain almost unchanged \cite{Fes58,Fes62,Mah69}.
Therefore the degree of overlap of neighboring 
resonances increases. This gives rise to interferences and the internal 
dynamics may suffer dramatic changes.
If the coupling strength passes over a critical value,
a segregation of the decay widths occurs.
Finally, for large couplings, $K$
rapidly decaying modes and $N-K$ narrow resonances 
result from this interference process \cite{Kle85,Sok89,Haa92}.
In the following we stick to the statistical approach \cite{Bro81,Por65}, 
which is more appropriate for the case of many resonances. Then the internal dynamics of
the system in consideration is modeled by a matrix ensemble. This may be chosen to describe
regular and chaotic motion respectively.

The distortion of {\it regular} motion by coupling a closed quantum system to the continuum 
has up to now received little attention theoretically \cite{Mue87,Dit91b,Seb96} and 
experimentally \cite{Sto90,Sto91,Mar93}. In contrast to that, such problems draw considerable 
attention in classical and semiclassical theories (e.g. \cite{Jun97} and references therein). 
It would be desirable to connect the approaches from both fields. The present work is ment to be
a first step along that way.

In \cite{Dit91b}, the regular motion  is described by a Poissonian Orthogonal
Ensemble (POE).  Its  perturbation due to its coupling to the continuum of
decay channels is investigated as a function of a parameter $\alpha$ which
characterizes the coupling strength.  
Level repulsion appears at  large $\alpha$ where the widths have already segregated.
There with increasing number of channels the correlations in the positions of the resonances
approach those, characteristic for the Gaussian Orthogonal Ensemble (GOE).
At the same time the distribution of the widths experiences a considerable
broadening in comparison to the $\chi^2_K$ distributed widths in the GOE case. This
broadening is quasi independent of the number of channels.
A similar effect has been found in the transition--strength dsitribution of a closed
system \cite{Alh86,Alh89} by increasing the chaoticity of the system. Note that a broad width
distribution implies deviations from the exponential decay law
\cite{Dit92,Per96}.

A physical example of a regular system, coupled to the continuum,
is investigated in \cite{Seb96,Sto91} on experimental and theoretical grounds.
The regular motion of the microwaves inside a rectangular resonator
is disturbed by an attached antenna, which defines the decay channel.
Theory predicts the effective coupling of the antenna to the resonator to increase
with the frequency.  The experimental data was obtained in the regime of strong external
coupling, where the resonance widths have already segregated.
It should be possible to apply the results obtained
in the present paper to this experimental setup.

In the following we compare the
results for a regular system described by a POE with those of a chaotic
system described by a GOE both coupled in the same way to the continuum.
Our interest focuses on the width
distribution and  the correlations induced  by the coupling of the  system to
the continuum of decay channels.

In Sec.\,2, we describe the statistical model used in our investigations as
well as some technicalities, concerning the numerical studies performed. 
A redistribution of the
spectroscopic properties takes place in the system if its coupling
 to the continuum of decay channels is sufficiently strong.  We sketch this
mechanism  ({\it trapping effect}) and give the formula for the mean width of
the long--lived resonances in Sec.\,3.   The distribution of the
widths  most characteristic of the trapping effect     is considered in
Sec.\,4.   In Sec.\,5, we present our results for the correlations
in the positions of the resonances while correlations
between widths and positions are discussed in Sec.\,6. Some conclusions
are drawn in the last section.

\section{Model \label{s_M}}

\subsection{Projection formalism \label{ss_for}}

Our analysis proceeds from the statistical model \cite{Mah69,Sok89}.
By using the projection operator technique \cite{Mah69} we
couple $N$ bound states $\; (|\Phi_i\rangle ,\;\;\; i=1,\ldots , N)$ ~to
$K$ common decay channels $\; (|\chi_c(E)\rangle ,\;\;\; c=1,\ldots , K)$
~via the two body residual interaction $V$.  We are
mainly interested in the bound state part of the system  and its
perturbations by the environment of decay channels.  To this end, we
 investigate the statistical properties of the system as a
function of the coupling strength  to the channels.

The total Hamiltonian of the whole quantum mechanical  system  consists
of three different parts: the Hamiltonian $H$ describing the
closed system of bound states, a term describing
the $K$ continua and a third term
which specifies the coupling between the two subspaces:
\bea
\label{Htot}
H_{\mbox{\tiny tot}} & = & \sum_{i,j=1}^{N} |\Phi_i\rangle
\; H_{ij} \;
 \langle\Phi_j|
 +  \sum_{c=1}^{K} \;\int dE\;\; |\chi_c(E)\rangle \; E   \;
\langle\chi_c(E)| \\
\nonumber
 &  & + \sum_{c=1}^{K}\sum_{i=1}^{N}\; \int dE\;\;  \left[
 \; |\Phi_i\rangle   \;
{\cal V}_i^c(E) \; \langle\chi_c(E)| + h.c. \right]       \; .
\eea
Here the ${\cal V}_i^c(E)$ are the coupling matrix elements
between the bound states $|\Phi_i\rangle$ and the scattering states $|\chi_c(E)\rangle$.
They can be understood as the
components of a $N$-dimensional vector ${\cal V}^c$. Its norm
is a measure of the coupling strength  to the channel $c$.
Direct reactions are neglected.  
We consider scattering systems with time reversal invariance, so we can restrict
ourselves to real matrices $H_{ij}$ and ${\cal V}_i^c(E)$.

As shown in \cite{Mah69} the S--matrix of such a system reads:
\begin{equation}
\label{Smat1}
S_{ab}(E)=\delta_{ab} \; - 2\pi{\rm i}\,  \sum_{i,j=1}^N
{\cal V}^a_i(E){\cal
G}_{ij}(E) {\cal V}^b_j(E)
\end{equation}
where ${\cal G}$ is the propagator in the perturbed system $H +
W^{\mbox{\tiny ext}}$,
\begin{equation}
\label{Prop}
{\cal G} =  \left[ E - H - W^{\mbox{\tiny ext}} \right]^{-1}   \; ,
\end{equation}
and the generally complex operator $W^{\mbox{\tiny ext}}$ describes
the perturbations from outside,
\begin{equation}
\label{Wext1}
W^{\mbox{\tiny ext}}_{ij} = \sum_{c=1}^K \;\int dE' \;\;\; \frac{{\cal V}_i^c(E')\;
 {\cal V}_j^c(E')}{E+i\eps-E'}.
\end{equation}
According to (\ref{Smat1}), the pole structure of the S--matrix is given by
the eigenvalues of the operator $H + W^{\mbox{\tiny ext}}$.
 We suppose, that the quantities ${\cal V}_i^c(E')$ are (almost) independent
 of the excitation energy, which is justified if one considers a finite
energy region sufficiently far from the reaction thresholds.  With this
assumption (\ref{Wext1}) simplifies
\begin{equation}
\label{Wext2}
W^{\mbox{\tiny ext}}_{ij} = -\frac{\rm i}{2}\,\eta \sum_{c=1}^K \; V_{ic} V_{jc}
\end{equation}
where we replaced ${\cal V}^c_i$ by $\sqrt{\frac{2\pi}{\eta}}V_{ic}$, 
$V_{ic}$ being a $N\times K$ matrix with on the average unit 
column vectors and $\eta$ being the total coupling strength per channel. 
Then the S--matrix takes the form
 \begin{equation}
\label{Smat2}
S(E) = {\bf 1}  - {\rm i}\eta \; V^+ \left[ E - \Heff
\right]^{-1} V \; .
\end{equation}
Its complex poles are given by the eigenvalues $\tilde{\cal
E}_n$ of the non--hermitian Hamiltonian
\begin{equation}
 \label{glHeff}
\Heff = H - \frac{\rm i}{2}\, \eta \; VV^+
 \end{equation}
with
\begin{equation}
\label{EW} \tilde{\cal E}_n = E_n - \frac{\rm i}{2} \Gamma_n \; .
\end{equation}

Under  the influence of the external coupling,
 the eigenstates of the hermitian matrix
$H$ are turning into resonances with a finite lifetime.  In the case of
non--overlapping resonances the real part $E_n$ and the imaginary
part $\Gamma_n$ in (\ref{EW})
give the position and the total decay width of the  resonance.
We will stick to these terms in all cases having in mind
the definition as poles of the $S$-matrix.

\subsection{Scattering ensemble \label{ss_SE}}

Following \cite{Sok89} we represent $\Heff$ in the eigenbasis of its hermitean part $H$.
Demanding orthogonal invariance of the scattering ensemble,
the matrix ensemble is characterized solely by the eigenvalue distribution of $H$,
and the anti--hermitean part $-\frac{{\rm i}\eta}{2} VV^+$ is independent on
the specific choice of the ensemble. In all cases $V$ consists of 
$K$ random column vectors. For large $N$ this implies the elements of $V$ being independent 
random Gaussian variables with
\be
 \langle V_{ia}\rangle = 0,\qquad \langle V_{ia}^2\rangle = \frac{1}{N} .
\ee

The pole distribution for the $\Heff$--ensemble may be expressed by
\cite{Sok89}:

\parbox{14.0cm}{
\begin{eqnarray*}
P(\{E_j,\Gamma_j\}) &=& C \int {\rm d}^N\{\eps_k\}\,{\rm d}^{NK}\{V_{la}\}\; 
\prod_j  \delta\left(
\det\left[{\T \Heff -E_j +{\frac{\rm i}{2}}\Gamma_j}\right]\right) \times\\
&&\prod_{m<n} |\eps_m-\eps_n|^\beta \exp\left(-{\T\frac{N}{a^2}}\sum_n\eps_n^2\right) \;
 \exp\left({-\T\frac{N}{2}}\sum_{n,a} V_{na}^2\right) .
\end{eqnarray*}}
\hfill \parbox{1cm}{\be\label{gl_jpd}\ee}

\noindent
$C$ is a normalization constant. The indices $j,k,l,m,n$ are running from $1\ldots N$ while
$a$ numbers the channels $1\ldots K$. The integration runs over the whole ensemble 
parametrised by the eigenvalues $\{\eps_k\}$ of $H$ and the coefficients $\{V_{la}\}$ of the
coupling matrix. By means of the $\delta$--functions, the positions
and widths of the resonances are introduced as the new variables of the distribution. 
The first part in the last line represents the eigenvalue
distribution of $H$, where $a$ is related to the range of the spectrum.
The last part in this line represents the distribution of the matrix elements
of $V$.

The parameter $\beta>-1$ controls the degree of level repulsion.
$\beta=1$ refers to the GOE and $\beta=0$ to the POE case.
$\beta\mapsto\infty$ describes a completely rigid spectrum (harmonic
oscillator) while $\beta < 0$ refers to a spectrum showing level clustering.
Note that the cases $\beta=2$ and $\beta=4$ do not describe the
unitary and symplectic ensembles, because we always work with real matrices $V$,
demanding {\it orthogonal invariance}.

In the following we restrict ourselves to the cases $\beta=1$ (GOE) and $\beta=0$ (POE).
For the GOE the level density is described by Wigner's semicircle \cite{Boh89},
$a$ being its radius. For the POE
(\ref{gl_jpd}) would produce a Gaussian shaped level density. 
But in order to compare the correlations in both ensembles we find it more convenient
to have equal level densities. Therefore we choose the $\{\eps_j\}$ in the POE case 
as distributed according to Wigner's semicircle law, too.

Unless stated otherwise, all numerical calculations were done by diagonalising
$\Heff$ of dimension $N=300$ and with $K=3$ channels being approximately orthogonal
to one another, due to the centralised distribution of their components. 
$V$ is a $N\times K$ matrix with random Gaussian coefficients with zero
mean and variance $1/N$. The calculations are performed
for $50$ matrices chosen from the GOE or the POE respectively.
Then the statistical observables are calculated as averages over
the central part of the spectrum and over the ensemble (the $50$ matrices) 
simultaneously. In the case of correlation measures ($\Sigma^2$ and $\Sigma_G^2$ in 
Sec.\,\ref{s_C} and Sec.\,\ref{s_CG}) the
positions of the resonances are unfolded to equal mean level spacing $d=1$.

\section{Resonance trapping  \label{s_T}}

\subsection{Separation of time scales}

One of the  specific properties of  open quantum systems is the
formation of separate time scales. In the strong coupling regime $\eta \gg 1$,
$K$ resonances become very broad by 'trapping' the remaining $N-K$ ones
\cite{Sok89,Rot91}.  Mathematically
the  trapping effect is caused by the fact that the rank of the hermitean part
of $\Heff$ is $N$ while that of the anti--hermitean part  $-\frac{{\rm i}\eta}{2}VV^+$
 cannot be larger than $K<N$ due to its  dyadic form (\ref{glHeff}). 
Thus we have the following picture:
  At  weak coupling the resonances do not overlap: $\langle \Gamma\rangle \ll D$,
where $D$ is the mean level distance, and the anti--hermitian part of $\Heff$ may be treated
by first order perturbation theory. Then the widths of the resonances
are well approximated by the diagonal elements of $VV^+$ and have comparable magnitude,
\begin{equation}
\Gamma_j =  \eta \sum_a V_{ia}^2 .
\label{widsm}
\end{equation}
In the opposite case of strong coupling, the resonances overlap and the
anti--hermitean part $-\frac{{\rm i}\eta}{2}VV^+$ 
dominates the behavior of the decaying system.
The appropriate basis is now the eigenbasis of $VV^+$ consisting of $K$
channel vectors (the columns of $V$) and $N-K$ vectors, spanning the kernel of
$VV^+$. The channel vectors have the common eigenvalue $\eta$.
The widths of the remaining states are zero.

In order to get some information about the magnitude of the widths
of the trapped resonances it is necessary to go beyond the simple Ansatz sketched
above.  This is done in the following subsection.

\subsection{Mean width of the trapped resonances}

We will calculate the mean width of the trapped resonances in the strong coupling regime
following \cite{Sok89,Sok92}. In addition to what has been done there, we show 
explicitely the applicability
to the POE case (in fact it may be applied to any orthogonal
invariant ensemble for arbitrary $\beta$ as defined in (\ref{gl_jpd}) ). In order to exploit
the orthogonal invariance we need the first and the second moments of the distributions of the
matrix elements in a generic (random) basis.
For the GOE they are \cite{Boh89}
\be
\langle H_{ij}\rangle = 0 ,\qquad 
\langle H_{ij}^2 \rangle = \left\{ \begin{array}{ccc}
\rule[-1mm]{0mm}{7mm}\frac{a^2}{2N} &:& i=j\\
\rule[-1mm]{0mm}{7mm}\frac{a^2}{4N} &:& i\ne j
\end{array} \right. .
\label{gl_GOEmel}\ee

For the POE we calculate these moments starting from the level distribution, and 
applying a random orthogonal transformation $O$:
\be 
H = O \diag\{\eps_j\} O^T,\qquad \langle O_{ij} \rangle = 0,\quad 
\langle O_{ij}^2\rangle = \frac{1}{N} .
\label{gl_raneig}\ee
So every element of $H$ is defined by
\be H_{ij} = \sum_k O_{ik} \eps_k O_{kj} .
\ee
Due to the centralised distribution of the $\{\eps_i\}$ the first moment 
$\langle H_{ij}\rangle$ is zero again. 
Then we calculate the average of the squares $H_{ij}^2$.  
As the eigenvalues are not correlated, we get
\be \langle H_{ij}^2\rangle =  \left\{ \begin{array}{ccc}
\rule[-1mm]{0mm}{7mm}\left\langle \sum_k O_{ik}^4 \eps_k^2 + 
\sum_{k\ne l} O_{ik}^2 O_{il}^2 \eps_k \eps_l \right\rangle =
{\frac{3}{N}} \langle \eps^2\rangle = \frac{3a^2}{4N}
&:& i=j \\
\rule[-1mm]{0mm}{7mm}\left\langle \sum_k O_{ik}^2 \eps_k^2 O_{jk}^2 \right\rangle =
{\frac{1}{N}} \langle \eps^2\rangle = 
\frac{a^2}{4N} 
&:& i\ne j
\end{array} \right. .
\label{gl_POEmel}\ee
Note that only the diagonal elements differ from the GOE case in their second moment.

In the following we show that the mean width of the trapped resonances can be calculated
from the second moments of the matrix elements $H_{ij}$.
To this end we turn to the channel representation of $\Heff$.
In this basis the channel vectors are the first $K$ canonical vectors and
the remaining $N-K$ vectors constitute a basis for the kernel of $VV^+$. Then
\be
\Heff = H \;-\; {\T\frac{\rm i}{2}}\eta \left( \begin{array}{cc} 
\delta_{ab} & 0 \\
0 & 0 
\end{array} \right) .
\ee
$a,b$ serve again as channel indices while the greek letters $\mu,\nu$ are used in connection
with the long--lived resonances. We diagonalise $H$ in kernel of $VV^+$ by the following 
orthogonal transformation
\be
P= \left( \begin{array}{cc}
\delta_{ab} & 0 \\
0 & P_{\mu\nu}
\end{array} \right) , \qquad D = P^T \Heff P = 
\left( \begin{array}{cc}
H_{ab} - {\T\frac{{\rm i}\eta}{2}} \delta_{ab} & * \\
  {*} & \begin{array}{ccc}
\eps_{K+1}' && 0 \\
 &\ddots& \\
 0 && \eps_N'
\end{array}
\end{array}
\right).
\ee
Note that $\{\eps_\nu'\}$ differ from the eigenvalues $\{\eps_j\}$ of $H$.
$D$ is symmetric and therefore the submatrices denoted by $*$ are transposed 
to each other,
\be
D_{\mu b} = \sum_{k,l=1}^N P_{k\mu} H_{kl} P_{lb} = \sum_{\nu=K+1}^N P_{\nu\mu} H_{\nu b} .
\ee
The $P_{\nu\mu}$ are only correlated with the matrix elements $H_{\mu\nu}$ as they are 
diagonalising this submatrix, but not with $H_{\mu b}$. Therefore one may average over 
these quantities independently.

Now we calculate approximate eigenvalues of $D$ by applying a Jacobi transformation of 
dimension $K+1$ \cite{Lor89}. Such a transformation is trace invariant 
(this is important, because the total width should be conserved). So
we look for the zero of the following determinant
(for arbitrary $K+1 \le\mu\le N$)
\be
 \det \left(\begin{array}{cc}
\tilde{\cal E}_\mu - \eps_\mu & D_{\mu 1} \ldots D_{\mu K} \\
\begin{array}{c}
D_{\mu 1} \\
\vdots \\
D_{\mu K}
\end{array} & \tilde{\cal E}_\mu -H_{ab} +{\T\frac{{\rm i}\eta}{2}} \delta_{ab}
\end{array} \right) = 0 .
\ee
Taking into account only the highest orders in $\eta$ we arrive at
\be
 \tilde{\cal E}_\mu -\eps_\mu + \sum_a \frac{D_{\mu a}^2}{{\cal E}- H_{aa} +\frac{{\rm i}\eta}{2}}
+{\cal O}(\eta^{-2}) = 0 .
\ee
We can read of the approximate position and width of a trapped resonance 
from the real and the imaginary part of $\tilde{\cal E}_\mu$. 
For $\eta\mapsto\infty$ we get
\be
E_\mu = \eps_\mu',\qquad \Gamma_\mu= {\frac{4}{\eta}} \sum_a D_{\mu a}^2.
\ee
The final step is the calculation of the ensemble mean of $D_{\mu a}^2$
(using (\ref{gl_GOEmel}) and (\ref{gl_POEmel}))
\be
 \langle D_{\mu a}^2\rangle = \langle \sum_\nu P_{\nu \mu}^2 H_{\nu a}^2 \rangle = 
\langle H_{\nu a}^2\rangle = \frac{\langle \eps^2\rangle}{N} .
\ee
From that it follows that the mean width for the trapped resonances is
\be
\langle \Gamma_\mu \rangle = {\frac{4}{\eta}} \langle \sum_a D_{\mu a}^2 \rangle =
\frac{4K}{N \eta} \langle \eps^2\rangle =
\frac{K a^2}{N \eta} .
\label{gl_gamf}\ee
Note that the diagonal elements of $H$ do not enter in the calculation of 
$\langle \Gamma_\mu\rangle$, so that (\ref{gl_gamf}) holds for both, the GOE and the POE.

Now one may consider $2\sqrt{\langle \eps^2\rangle}$ as the 'width' of the level density and
then introduce a dimensionless parameter $\kappa$ measuring the degree of overlapping:
The summed width per channel divided by the 'width' of the spectrum,
\be
\kappa = \frac{\eta}{2\sqrt{\langle\eps^2\rangle}} .
\label{gl_dkap} 
\ee
For the trapped resonances this overlapping parameter is
\be
\kappa_f = \frac{(N-K)\langle \Gamma_\mu\rangle}{2K\sqrt{\langle\eps^2\rangle}}.
\ee
Then it follows from (\ref{gl_gamf}) that in the limit $N\mapsto\infty$
\be
\kappa_f = \frac{N-K}{N}\, \frac{2\sqrt{\langle\eps^2\rangle}}{\eta} \mapsto \kappa^{-1} .
\ee

Assuming ergodicity (as expressed by the randomness of the eigenbasis (\ref{gl_raneig}) )
the ensemble average may be replaced by the spectral average.
Then (\ref{gl_gamf}) may even be applied to a
single system. In this case it may serve as a test for the trapping scenario
to occur in the system considered: (\ref{gl_gamf}) relates the product of the average width 
$\langle\Gamma_\mu\rangle$ of the trapped resonances and the average width $\eta$ of the broad 
resonances to the variance of the level density $\langle \eps^2\rangle$.

For small $\kappa$ one can still consider the mean width of the $N-K$ smallest resonances
in order to have $\kappa_f$ well defined globally. Then this quantity
will be equal to the mean width of all resonances, if $\kappa$ is sufficiently small. Therefore:
\be
\kappa_f = \left\{ \begin{array}{rcl}
\kappa &:& \kappa\ll 1\\
\kappa^{-1} &:& \kappa\gg 1
\end{array} \right. .
\label{gl_kap}
\ee
We define the weak coupling regime by $\kappa \ll 1$, so that the upper part of (\ref{gl_kap})
is fulfilled, and the strong coupling regime by $\kappa \gg 1$ so that the lower part becomes valid.
In between (critical region) the redistribution takes place. 

For illustration, in Fig.\,1
 the trapping process is portrayed for both the GOE and the POE,
from the low coupling to the high coupling regime. We plot $\kappa_f$
versus $\kappa$ as defined in (\ref{gl_dkap}).  A double--log plot is used in
order to demonstrate the proportionality relation (\ref{gl_kap}).  
As the level density is the same for the GOE and the POE (cf. Subsec.\,\ref{ss_SE}),
the data points of both the GOE and the POE case follow
the same line for all coupling strengths within the numerical accuracy.
The asymmetry on the left side of the plot is due to the finite dimension
$N=300$ in our calculations.

The redistribution between the two scenarios at small and large
coupling  $\eta$ occurs rather promptly at  $\kappa \approx 1$. Here the  
$K$ broad poles appear which will share almost the total sum of the widths, 
whereas the remaining $N-K$ resonances will become more and more long--lived in the strong 
coupling regime. 
The point $\kappa=1$, at which $\kappa_f$ reaches its maximal value, is called
critical point. 
Note the peculiar properties of the transmission coefficient \cite{Ver85,Mue87} and
of the width distribution \cite{Fyo96} at this point.

\begin{figure}[f]
\begin{picture}(0,0)%
\epsfbox{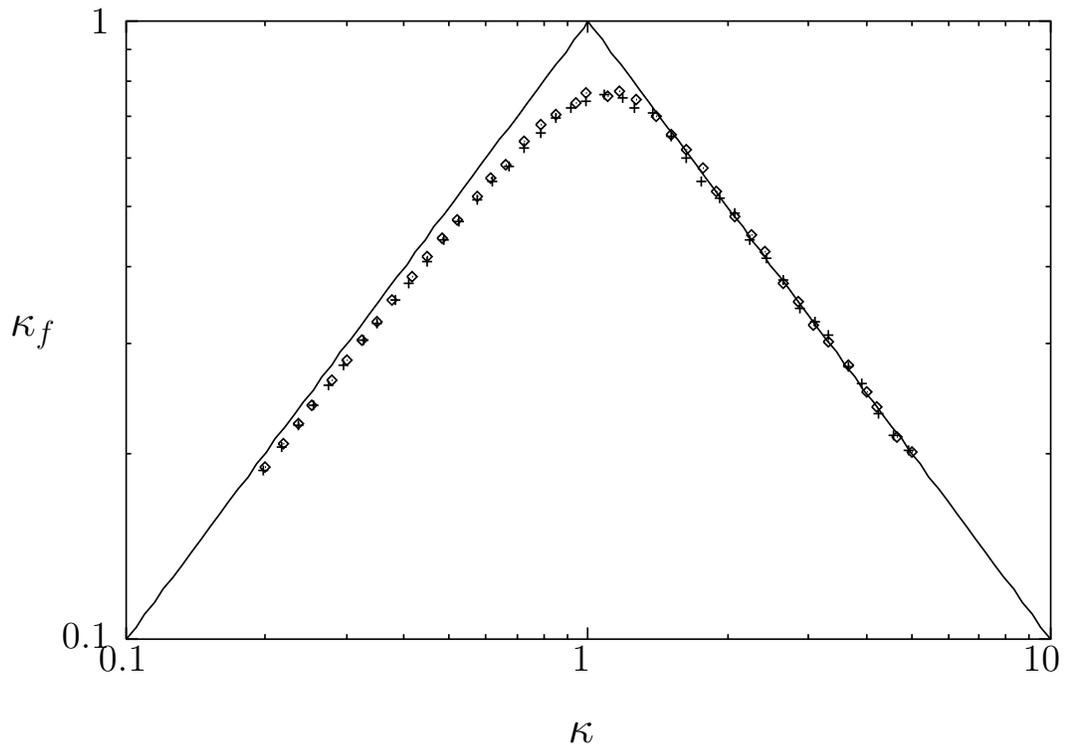}%
\end{picture}%
\setlength{\unitlength}{0.00058300in}%
\begingroup\makeatletter\ifx\SetFigFont\undefined
\def\x#1#2#3#4#5#6#7\relax{\def\x{#1#2#3#4#5#6}}%
\expandafter\x\fmtname xxxxxx\relax \def\y{splain}%
\ifx\x\y   
\gdef\SetFigFont#1#2#3{%
  \ifnum #1<17\tiny\else \ifnum #1<20\small\else
  \ifnum #1<24\normalsize\else \ifnum #1<29\large\else
  \ifnum #1<34\Large\else \ifnum #1<41\LARGE\else
     \huge\fi\fi\fi\fi\fi\fi
  \csname #3\endcsname}%
\else
\gdef\SetFigFont#1#2#3{\begingroup
  \count@#1\relax \ifnum 25<\count@\count@25\fi
  \def\x{\endgroup\@setsize\SetFigFont{#2pt}}%
  \expandafter\x
    \csname \romannumeral\the\count@ pt\expandafter\endcsname
    \csname @\romannumeral\the\count@ pt\endcsname
  \csname #3\endcsname}%
\fi
\fi\endgroup
\begin{picture}(10695,6636)(139,-6436)
\put(10696,-5716){\makebox(0,0)[b]{\smash{\SetFigFont{14}{16.8}{rm}10}}}
\put(1801,-2686){\makebox(0,0)[rb]{\smash{\SetFigFont{17}{20.4}{rm}$\kappa_f$}}}
\put(6541,-6361){\makebox(0,0)[b]{\smash{\SetFigFont{17}{20.4}{rm}$\kappa$}}}
\put(2281, 44){\makebox(0,0)[rb]{\smash{\SetFigFont{14}{16.8}{rm}1}}}
\put(2281,-5506){\makebox(0,0)[rb]{\smash{\SetFigFont{14}{16.8}{rm}0.1}}}
\put(2401,-5716){\makebox(0,0)[b]{\smash{\SetFigFont{14}{16.8}{rm}0.1}}}
\put(6541,-5716){\makebox(0,0)[b]{\smash{\SetFigFont{14}{16.8}{rm}1}}}
\end{picture}
\caption{The overlapping parameter $\kappa_f$ of the trapped resonances versus the overlapping
parameter $\kappa$ of all resonances. The diamonds denote data points for the GOE, the crosses
those for the POE. The solid lines show the two asymptotics $\kappa$ and 
$\kappa^{-1}$.}
\label{f_tra}
\end{figure}

\section{Width distribution \label{s_W}}

In the weak coupling  regime the widths are approximately given by
the diagonal elements of the coupling matrix $VV^+$ (\ref{widsm}).
These are sums of $K$ random  Gaussian variables and therefore $\chi^2_K$
distributed when normalised to unit mean
\be
p(y) = \chi^2_K(y) = \frac{(K/2)^{\frac{K}{2}}}{\Gamma(K/2)} \,
y^{\frac{K}{2} -1} \,{\rm e}^{-\frac{K}{2} y}.
\label{gl4}
\ee
For the GOE the same distribution holds in the  
strong coupling region as shown for the one channel case in \cite{Sok89}, by
calculating the joint probability distribution for the complex eigenvalues of
$\Heff$ explicitely.  For the POE the width distribution becomes much broader.

The numerical results are shown in 
Fig.\,2. Here a series of width distributions for the GOE (left side) and
the POE (right side) for increasing coupling strength $\kappa$ (from top to
bottom) is given.  When $\kappa > 1$, only the long--lived resonances  were
taken into account. In each diagram, we plotted the numerical data
(histogram), and the $\chi^2_{K=3}$ distribution (dashed line). For the POE
at $\kappa=100$ a best fit $\chi^2_{q}$ distribution (dotted line) is given. 
This distribution is calculated by replacing in (\ref{gl4}) $K$ by a real parameter 
$q$ and performing a $\chi^2$ fit in order to find the best estimate for $q$.

For both ensembles the GOE and the POE, the width distributions undergo strong 
deformations in the critical region, but become stationary again in the strong coupling
limit. In contrast to the GOE case, which returns
to its old shape, the distribution for the POE remains much
broader. It is in good agreement with the best fit $\chi_q^2$
distribution with $q=1.7$.

\begin{figure}[f]
\begin{picture}(0,0)%
\epsfbox{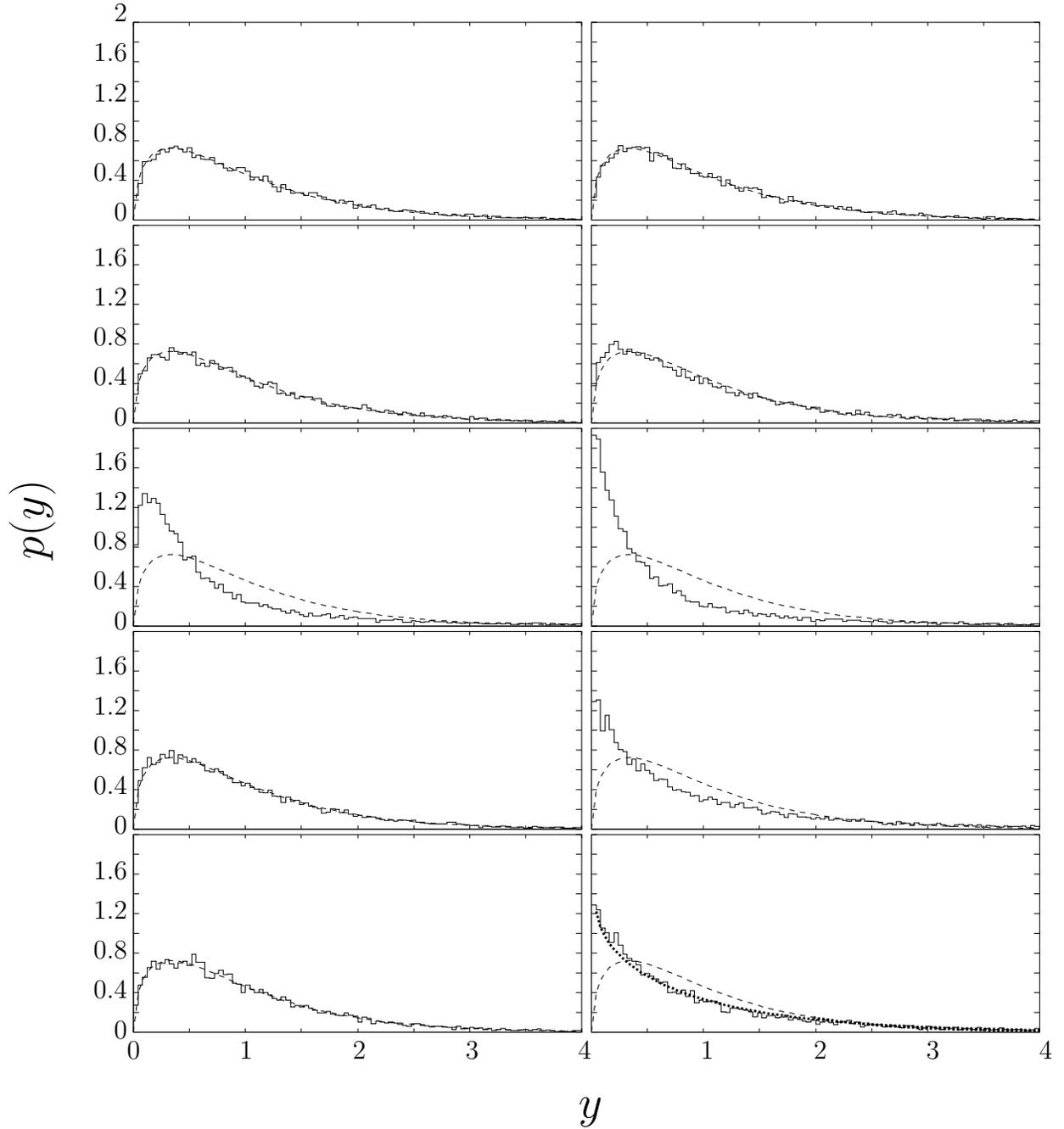}%
\end{picture}%
\setlength{\unitlength}{0.00033300in}%
\begingroup\makeatletter\ifx\SetFigFont\undefined
\def\x#1#2#3#4#5#6#7\relax{\def\x{#1#2#3#4#5#6}}%
\expandafter\x\fmtname xxxxxx\relax \def\y{splain}%
\ifx\x\y   
\gdef\SetFigFont#1#2#3{%
  \ifnum #1<17\tiny\else \ifnum #1<20\small\else
  \ifnum #1<24\normalsize\else \ifnum #1<29\large\else
  \ifnum #1<34\Large\else \ifnum #1<41\LARGE\else
     \huge\fi\fi\fi\fi\fi\fi
  \csname #3\endcsname}%
\else
\gdef\SetFigFont#1#2#3{\begingroup
  \count@#1\relax \ifnum 25<\count@\count@25\fi
  \def\x{\endgroup\@setsize\SetFigFont{#2pt}}%
  \expandafter\x
    \csname \romannumeral\the\count@ pt\expandafter\endcsname
    \csname @\romannumeral\the\count@ pt\endcsname
  \csname #3\endcsname}%
\fi
\fi\endgroup
\begin{picture}(19115,20559)(763,-20383)
\put(1351,-9961){\makebox(0,0)[lb]{\smash{
\SetFigFont{20}{24.0}{rm}$p(y)$
}}}
\put(2881,-18775){\makebox(0,0)[rb]{\smash{\SetFigFont{12}{14.4}{rm}0}}}
\put(2881,-18042){\makebox(0,0)[rb]{\smash{\SetFigFont{12}{14.4}{rm}0.4}}}
\put(2881,-17310){\makebox(0,0)[rb]{\smash{\SetFigFont{12}{14.4}{rm}0.8}}}
\put(2881,-16577){\makebox(0,0)[rb]{\smash{\SetFigFont{12}{14.4}{rm}1.2}}}
\put(2881,-15844){\makebox(0,0)[rb]{\smash{\SetFigFont{12}{14.4}{rm}1.6}}}
\put(2881,-15013){\makebox(0,0)[rb]{\smash{\SetFigFont{12}{14.4}{rm}0}}}
\put(2881,-14280){\makebox(0,0)[rb]{\smash{\SetFigFont{12}{14.4}{rm}0.4}}}
\put(2881,-13548){\makebox(0,0)[rb]{\smash{\SetFigFont{12}{14.4}{rm}0.8}}}
\put(2881,-12815){\makebox(0,0)[rb]{\smash{\SetFigFont{12}{14.4}{rm}1.2}}}
\put(2881,-12082){\makebox(0,0)[rb]{\smash{\SetFigFont{12}{14.4}{rm}1.6}}}
\put(2881,-11251){\makebox(0,0)[rb]{\smash{\SetFigFont{12}{14.4}{rm}0}}}
\put(2881,-10518){\makebox(0,0)[rb]{\smash{\SetFigFont{12}{14.4}{rm}0.4}}}
\put(2881,-9786){\makebox(0,0)[rb]{\smash{\SetFigFont{12}{14.4}{rm}0.8}}}
\put(2881,-9053){\makebox(0,0)[rb]{\smash{\SetFigFont{12}{14.4}{rm}1.2}}}
\put(2881,-8320){\makebox(0,0)[rb]{\smash{\SetFigFont{12}{14.4}{rm}1.6}}}
\put(2881,-7489){\makebox(0,0)[rb]{\smash{\SetFigFont{12}{14.4}{rm}0}}}
\put(2881,-6756){\makebox(0,0)[rb]{\smash{\SetFigFont{12}{14.4}{rm}0.4}}}
\put(2881,-6024){\makebox(0,0)[rb]{\smash{\SetFigFont{12}{14.4}{rm}0.8}}}
\put(2881,-5291){\makebox(0,0)[rb]{\smash{\SetFigFont{12}{14.4}{rm}1.2}}}
\put(2881,-4558){\makebox(0,0)[rb]{\smash{\SetFigFont{12}{14.4}{rm}1.6}}}
\put(13546,-19156){\makebox(0,0)[b]{\smash{\SetFigFont{12}{14.4}{rm}1}}}
\put(15631,-19156){\makebox(0,0)[b]{\smash{\SetFigFont{12}{14.4}{rm}2}}}
\put(17701,-19156){\makebox(0,0)[b]{\smash{\SetFigFont{12}{14.4}{rm}3}}}
\put(19771,-19156){\makebox(0,0)[b]{\smash{\SetFigFont{12}{14.4}{rm}4}}}
\put(3001,-19156){\makebox(0,0)[b]{\smash{\SetFigFont{12}{14.4}{rm}0}}}
\put(7156,-19156){\makebox(0,0)[b]{\smash{\SetFigFont{12}{14.4}{rm}2}}}
\put(9226,-19156){\makebox(0,0)[b]{\smash{\SetFigFont{12}{14.4}{rm}3}}}
\put(11296,-19156){\makebox(0,0)[b]{\smash{\SetFigFont{12}{14.4}{rm}4}}}
\put(5071,-19156){\makebox(0,0)[b]{\smash{\SetFigFont{12}{14.4}{rm}1}}}
\put(2881,-3727){\makebox(0,0)[rb]{\smash{\SetFigFont{12}{14.4}{rm}0}}}
\put(2881,-2994){\makebox(0,0)[rb]{\smash{\SetFigFont{12}{14.4}{rm}0.4}}}
\put(2881,-2262){\makebox(0,0)[rb]{\smash{\SetFigFont{12}{14.4}{rm}0.8}}}
\put(2881,-1529){\makebox(0,0)[rb]{\smash{\SetFigFont{12}{14.4}{rm}1.2}}}
\put(2881,-796){\makebox(0,0)[rb]{\smash{\SetFigFont{12}{14.4}{rm}1.6}}}
\put(2881,-64){\makebox(0,0)[rb]{\smash{\SetFigFont{12}{14.4}{rm}2}}}
\put(11401,-20236){\makebox(0,0)[b]{\smash{\SetFigFont{20}{24.0}{rm}$y$}}}
\end{picture}
\caption{Normalized width distributions for the GOE (left side) and POE
(right side) for different coupling strengths $\kappa=0.01, 0.1, 1, 10, 100$ 
(from top to bottom). Histogram of the width distribution obtained numerically
(solid line). $\chi^2_{K=3}$ distribution (dashed line). Best fit $\chi^2_{q=1.7}$
distribution (dotted line) on the bottom left. }
\label{widis}
\end{figure}

Further investigations of the POE at strong coupling for more channels led to the
following interesting behavior (cf. Fig.\,3):
The coupling to continuum leads to a broadening of the width distribution.
For increasing number of channels $K$
its variance approaches double what it was at small coupling (or 
double what it is in the corresponding GOE case).
Whereas the variances of the best fit 
distributions systematically underestimate the factor $2$, the numerical
variances overestimate it. 

\begin{figure}[f]
\begin{picture}(0,0)%
\epsfbox{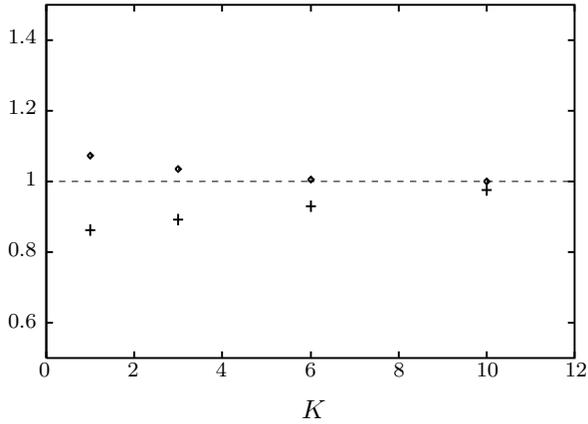}%
\end{picture}%
\setlength{\unitlength}{0.00033300in}%
\begingroup\makeatletter\ifx\SetFigFont\undefined
\def\x#1#2#3#4#5#6#7\relax{\def\x{#1#2#3#4#5#6}}%
\expandafter\x\fmtname xxxxxx\relax \def\y{splain}%
\ifx\x\y   
\gdef\SetFigFont#1#2#3{%
  \ifnum #1<17\tiny\else \ifnum #1<20\small\else
  \ifnum #1<24\normalsize\else \ifnum #1<29\large\else
  \ifnum #1<34\Large\else \ifnum #1<41\LARGE\else
     \huge\fi\fi\fi\fi\fi\fi
  \csname #3\endcsname}%
\else
\gdef\SetFigFont#1#2#3{\begingroup
  \count@#1\relax \ifnum 25<\count@\count@25\fi
  \def\x{\endgroup\@setsize\SetFigFont{#2pt}}%
  \expandafter\x
    \csname \romannumeral\the\count@ pt\expandafter\endcsname
    \csname @\romannumeral\the\count@ pt\endcsname
  \csname #3\endcsname}%
\fi
\fi\endgroup
\begin{picture}(8892,6561)(1117,-6424)
\put(9871,-5716){\makebox(0,0)[b]{\smash{\SetFigFont{8}{9.6}{rm}12}}}
\put(5776,-6361){\makebox(0,0)[b]{\smash{\SetFigFont{10}{12.0}{rm}$K$}}}
\put(1456,-4951){\makebox(0,0)[rb]{\smash{\SetFigFont{8}{9.6}{rm}0.6}}}
\put(1456,-3841){\makebox(0,0)[rb]{\smash{\SetFigFont{8}{9.6}{rm}0.8}}}
\put(1456,-2731){\makebox(0,0)[rb]{\smash{\SetFigFont{8}{9.6}{rm}1}}}
\put(1456,-1621){\makebox(0,0)[rb]{\smash{\SetFigFont{8}{9.6}{rm}1.2}}}
\put(1456,-511){\makebox(0,0)[rb]{\smash{\SetFigFont{8}{9.6}{rm}1.4}}}
\put(1576,-5716){\makebox(0,0)[b]{\smash{\SetFigFont{8}{9.6}{rm}0}}}
\put(2956,-5716){\makebox(0,0)[b]{\smash{\SetFigFont{8}{9.6}{rm}2}}}
\put(4336,-5716){\makebox(0,0)[b]{\smash{\SetFigFont{8}{9.6}{rm}4}}}
\put(5731,-5716){\makebox(0,0)[b]{\smash{\SetFigFont{8}{9.6}{rm}6}}}
\put(7111,-5716){\makebox(0,0)[b]{\smash{\SetFigFont{8}{9.6}{rm}8}}}
\put(8491,-5716){\makebox(0,0)[b]{\smash{\SetFigFont{8}{9.6}{rm}10}}}
\end{picture}
\caption{Relative deviations of the variance of the width distributions 
(POE, strong coupling) for $K= 1, 3, 6, 10$ channels. The diamonds denote the
variances calculated directly from the data, whereas the crosses denote $2/q$, the
variances of the best fit $\chi_q^2$ distributions. These values were divided by $4/K$,
which seems to be the limit value for the variance at $K\mapsto\infty$. }
\label{table}
\end{figure}

In order to understand the broadening of the width distributions due to the coupling to
continuum, we consider the one channel case in more detail. The S--matrix (\ref{Smat2}) with 
complex argument ${\cal E}$ can equivalently be written as \cite{Sok92}
\be
S({\cal E}) = \frac{1-{\rm i}R}{1+{\rm i}R},\qquad 
R= \frac{\eta}{2} \,V^+ \frac{1}{{\cal E}-H} V .
\ee
From that it follows that the poles of the S--matrix are given
by the zeros of the function
\be 
f({\cal E}) = 1 + \frac{\rm i}{2}\, \eta \sum_i \frac{v_i^2}{{\cal E} -\eps_i} .
\ee
The $v_i$ are the Gaussian random coefficients of the coupling matrix (\ref{widsm}) with 
variance $\frac{1}{N}$ and the $\eps_i$ are the real eigenvalues of $H$. 
We may consider the real and the imaginary part of $f$ separately. As we are interested in the
case $\eta\mapsto\infty$ we only keep the
highest order of $\eta$, making use of $\Gamma \sim \eta^{-1}$ (\ref{gl_gamf}). Using the
notation ${\cal E} = E -\frac{\rm i}{2}\Gamma$ for the poles one arrives at \cite{Seb96}
\be \sum_i \frac{v_i^2}{E-\eps_i} = 0,\qquad 
\frac{4}{\eta\Gamma} = \sum_i \frac{v_i^2}{(E-\eps_i)^2} .
\label{gl_polesc}
\ee
Between every two neighboring levels there has to settle down one pole with increasing $\eta$ 
due to the structure of (\ref{gl_polesc}). Choosing an arbitrary one of them
and taking into account only its two neighboring levels, we find the following formula 
for its width (a detailed derivation is given in App.\,\ref{as_W}):
\be \Gamma =  \frac{d_0^2\, s^2}{\eta(v_1^2+v_2^2)} (1-\tau^2),\qquad
\tau= \frac{v_2^2-v_1^2}{v_1^2+v_2^2} .
\label{gl_gdis1}
\ee
Here $s$ is the distance between the two consecutive levels measured in units of the
mean level distance $d_0$ in the center of the spectrum.  In our case of a semicircular
level density with radius $a$, $d_0 = \frac{\pi a}{2 N}$.  
In App.\,\ref{as_WD} we calculate the distribution of the trapped widths normalised by their mean 
$\langle \Gamma_\nu\rangle$ (\ref{gl_gamf}) for the GOE and the POE case. This is done
by evaluating the following integral:
\be
p \left(y= \Gamma/\langle \Gamma_\nu\rangle \right) =
\frac{N}{2\pi} \int \d s\, P(s)\;\int \d v_1\,\d v_2\, 
{\rm e}^{\T -\frac{N}{2}(v_1^2+v_2^2)}\;
\delta\left[ y- \frac{\pi^2\, s^2}{4N} \frac{1-\tau^2}{v_1^2+v_2^2} \right] .
\ee
Here $P(s)$ is the nearest neighbor distribution for the ensemble in consideration. The result
for the GOE is:
\be p_{\rm G}(y) = \frac{1}{\sqrt{2\pi y}}\, \left( 1+ \frac{2y}{\pi} \right)^{-3/2}
\label{gl_gk1wG}
\ee
and for the POE it is
\be
p_{\rm P}(y) = \frac{1}{4y}\, {\rm e}^{\T\frac{y}{\pi^2}}\,
{\rm W}_{-1,0}\left(\T\frac{2y}{\pi^2}\right) ,
\label{gl_gk1wP}
\ee
where
\be
{\rm W}_{-1,0}(z) = \frac{2z}{\sqrt{\pi}}\left[ \left(\T\sqrt{z}+\frac{\T 1}{\T \sqrt{z}}\right) 
K_0(\T\frac{z}{2}) -\sqrt{z} K_1(\T\frac{z}{2}) \right] 
\ee
is the Whittaker function \cite{GRy80} and $K_0(z)$ and $K_1(z)$ are the modified Bessel functions 
\cite{ASt64}. In Fig.\,4 both distributions are compared to the Porter Thomas 
distribution $\chi^2_{K=1}$ (\ref{gl4}) and the numerically obtained width distribution for 
the POE case. Note that we plotted $p( \ln y)$ instead of $p(y)$ for all distributions in
this figure, and we use a logarithmic scale on the abscissa.
This is done in order to give an overall view and allowing in the same time for the 
recognition of the interesting features discussed in the following.

Comparing (\ref{gl_gk1wG}) to the Porter Thomas distribution, we find that for
$y\ll 1$ it agrees exactly  in the leading power.
For large $y\gg 1$ $p_{\rm G}$ fails: it has a $y^{-2}$ tail which actually makes the
distribution non normalisable.

For the POE case we observe the same features when comparing (\ref{gl_gk1wP})
to the width distribution obtained numerically. For $y\to\infty$,
$W_{-1,0}(z) \to \exp(-z/2)/z$ leading again to a $y^{-2}$ tail being inconsistent with the
numerical result.  On the other hand for $y\to 0$, $p_{\rm P}(y)$ fits very well to the numerical
distribution. Here $K_1(z) \to 1/z$ and $K_0(z)\to -{\rm ln}z$
and therefore
\be
p_{\rm P}(y) \to -\frac{\ln y}{\pi\sqrt{2\pi y}} .
\ee

In the relevant range $y\approx 1$ both distributions $p_{\rm G}(y)$ and $p_{\rm P}(y)$ 
show qualitatively how the level repulsion parameter  $\beta$ (\ref{gl_jpd}) affects
the width distribution of the trapped resonances. 
Disregarding the tails, one can clearly see from Fig.\,4 the broadening in the
POE case ($\beta =0$) in comparisson to the GOE ($\beta =1$).

\begin{figure}[f]
\begin{picture}(0,0)%
\epsfbox{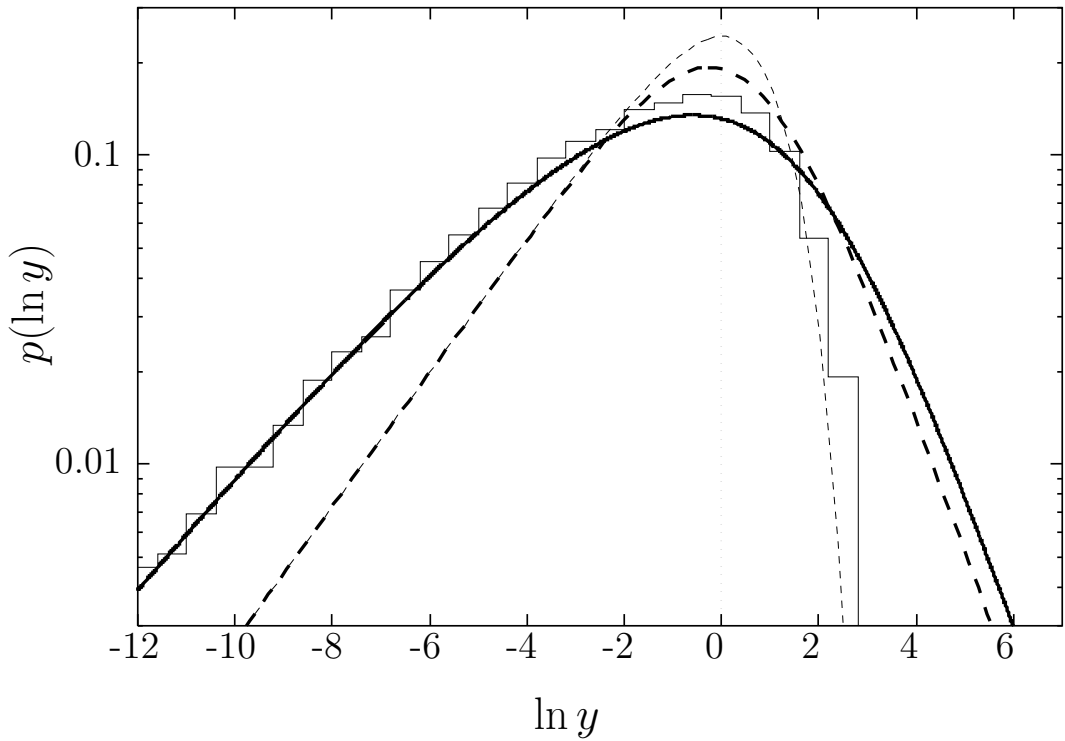}%
\end{picture}%
\setlength{\unitlength}{0.00058300in}%
\begingroup\makeatletter\ifx\SetFigFont\undefined
\def\x#1#2#3#4#5#6#7\relax{\def\x{#1#2#3#4#5#6}}%
\expandafter\x\fmtname xxxxxx\relax \def\y{splain}%
\ifx\x\y   
\gdef\SetFigFont#1#2#3{%
  \ifnum #1<17\tiny\else \ifnum #1<20\small\else
  \ifnum #1<24\normalsize\else \ifnum #1<29\large\else
  \ifnum #1<34\Large\else \ifnum #1<41\LARGE\else
     \huge\fi\fi\fi\fi\fi\fi
  \csname #3\endcsname}%
\else
\gdef\SetFigFont#1#2#3{\begingroup
  \count@#1\relax \ifnum 25<\count@\count@25\fi
  \def\x{\endgroup\@setsize\SetFigFont{#2pt}}%
  \expandafter\x
    \csname \romannumeral\the\count@ pt\expandafter\endcsname
    \csname @\romannumeral\the\count@ pt\endcsname
  \csname #3\endcsname}%
\fi
\fi\endgroup
\begin{picture}(9442,6562)(451,-6436)
\put(2446,-5716){\makebox(0,0)[b]{\smash{\SetFigFont{14}{16.8}{rm}-10}}}
\put(3316,-5716){\makebox(0,0)[b]{\smash{\SetFigFont{14}{16.8}{rm}-8}}}
\put(4201,-5716){\makebox(0,0)[b]{\smash{\SetFigFont{14}{16.8}{rm}-6}}}
\put(5071,-5716){\makebox(0,0)[b]{\smash{\SetFigFont{14}{16.8}{rm}-4}}}
\put(5941,-5716){\makebox(0,0)[b]{\smash{\SetFigFont{14}{16.8}{rm}-2}}}
\put(6811,-5716){\makebox(0,0)[b]{\smash{\SetFigFont{14}{16.8}{rm}0}}}
\put(7681,-5716){\makebox(0,0)[b]{\smash{\SetFigFont{14}{16.8}{rm}2}}}
\put(8566,-5716){\makebox(0,0)[b]{\smash{\SetFigFont{14}{16.8}{rm}4}}}
\put(9436,-5716){\makebox(0,0)[b]{\smash{\SetFigFont{14}{16.8}{rm}6}}}
\put(1576,-5716){\makebox(0,0)[b]{\smash{\SetFigFont{14}{16.8}{rm}-12}}}
\put(751,-3061){\makebox(0,0)[lb]{\smash{
\SetFigFont{17}{20.4}{rm}$p(\ln y)$
}}}
\put(1456,-4051){\makebox(0,0)[rb]{\smash{\SetFigFont{14}{16.8}{rm}0.01}}}
\put(1456,-1276){\makebox(0,0)[rb]{\smash{\SetFigFont{14}{16.8}{rm}0.1}}}
\put(5476,-6361){\makebox(0,0)[b]{\smash{\SetFigFont{17}{20.4}{rm}$\ln y$}}}
\end{picture}
\caption{Distributions of the logarithms of the widths in the strong coupling limit for 
the one channel 
case. The prediction of the two level approximation for the POE (\ref{gl_gk1wP})
(solid bold line) is compared to the numerical result (solid line) produced by 
diagonalising $200$ matrices at $\kappa=100$. The corresponding two level GOE result
(\ref{gl_gk1wG}) (dashed bold line) is compared to the exact distribution, the Porter 
Thomas curve (dashed line).}
\label{f_k1wid}
\end{figure}

\section{Correlations in the positions of the resonances \label{s_C}}

In order to measure the correlations between the positions of the resonances,
we apply the number variance $\Sigma^2$ \cite{Boh89}. We investigate the GOE and the POE
spectra in the different coupling regimes: weak, critical and strong coupling. For the
GOE similar investigations have already been done in \cite{Miz93}.

$\Sigma^2$ measures the deviation of the accumulated level density from a straight 
line. Small $\Sigma^2$ is a signature for high {\it rigidity}
of the level sequence. In this case the level spacings are more or less
equal. On the other hand a completely uncorrelated sequence, as for example the POE
spectrum has minimal rigidity and therefore maximal $\Sigma^2$.

In contrast to \cite{Dit91b} we focus on the few channel case (typically $K=3$), because
we are mainly interested in the differences the GOE and the POE show at large coupling.
Note that one result of \cite{Dit91b} was that the POE spectrum at strong coupling becomes more
and more GOE--like with increasing number of channels.

For the numerical calculations we implemented the following technical steps: The spectra
are unfolded to constant mean level spacing $d=1$. This is done by a polynomial fit of the
accumulated level density.
Then the edges of the spectrum are skipped, which reduces the number of resonances by
approximately $25\%$. Furthermore when $\kappa > 1$, the $K$ broadest resonances are 
omitted.  For the new reduced spectrum $\{e_i\}$ consisting now of $N'$ levels the correlations 
are investigated. 

For an unfolded sequence $\{e_i\}$, the mean number of levels found in an interval of 
length $l$ is $\langle n(l)\rangle = l$ due to $d=1$. The number variance $\Sigma^2(l)$
is defined as the variance of $n(l)$ in the ensemble mean:
\be 
\Sigma^2(l) = \langle n^2(l)\rangle - \langle n(l)\rangle^2  =
\langle n^2(l)\rangle - l^2 .
\label{gl_S2l}
\ee
For the GOE and the POE (completely uncorrelated sequence) this quantity is 
known analytically \cite{Boh89}. Instead for our purposes it is sufficient to use 
the following approximate expressions:

For the GOE
\be
\Sigma^2_{\rm G}(l) \approx \frac{2 \ln l}{\pi^2} +0.442
\label{gl19}\label{gl_S2G}
\ee

and for the POE
\be
\Sigma^2_{\rm P}(l) \approx l .
\label{gl_S2P}\ee

In Fig.\,5 we show numerical calculations of $\Sigma^2$ for the GOE (a)
and the POE (b) spectra. This is done for three different values of the overlapping 
parameter $\kappa= 0.1, 1$, and $10$. The smallest and the largest value are chosen such
that a further decrease or increase of $\kappa$ does not change any more the outcome for 
$\Sigma^2$ significantly.

$\Sigma^2$ for the GOE (Fig.\,5a) shows good agreement with $\Sigma^2_{\rm G}$
(\ref{gl19}) for $\kappa=0.1$ and $\kappa=10$. Only in the critical region we notice a 
deviation, where $\Sigma^2$ becomes slightly larger for all $l>1$. This decorrelation
is due to the additional 'degree of freedom' the resonance poles encounter in this
regime \cite{Mue95}. In contrast to the two other cases, where the poles are restricted 
to a small stripe along the real axis, they now have enough space in the direction of 
the imaginary axis in order to avoid close neighbors. This 'width repulsion' cannot be
detected by observing the positions alone.

In the POE case (Fig.\,5b) all correlations are induced by the coupling to the decay channels.
Therefore they grow with increasing $\kappa$. After completion of the redistribution
(i.e. when the trapped resonances come sufficiently close to the real axis again) the
$\Sigma^2$ curve becomes stable. This happens at $\kappa=10$. Note that
the deviations from the original straight line do not
vanish nor does the curve at large $\kappa$ coincide with that of the GOE.  
Furthermore it is remarkable, that at $\kappa <1$ the
rigidity increases mainly for large $l$ and catches up at $\kappa >1$
for small $l$.

\begin{figure}[f]
\begin{picture}(0,0)%
\epsfbox{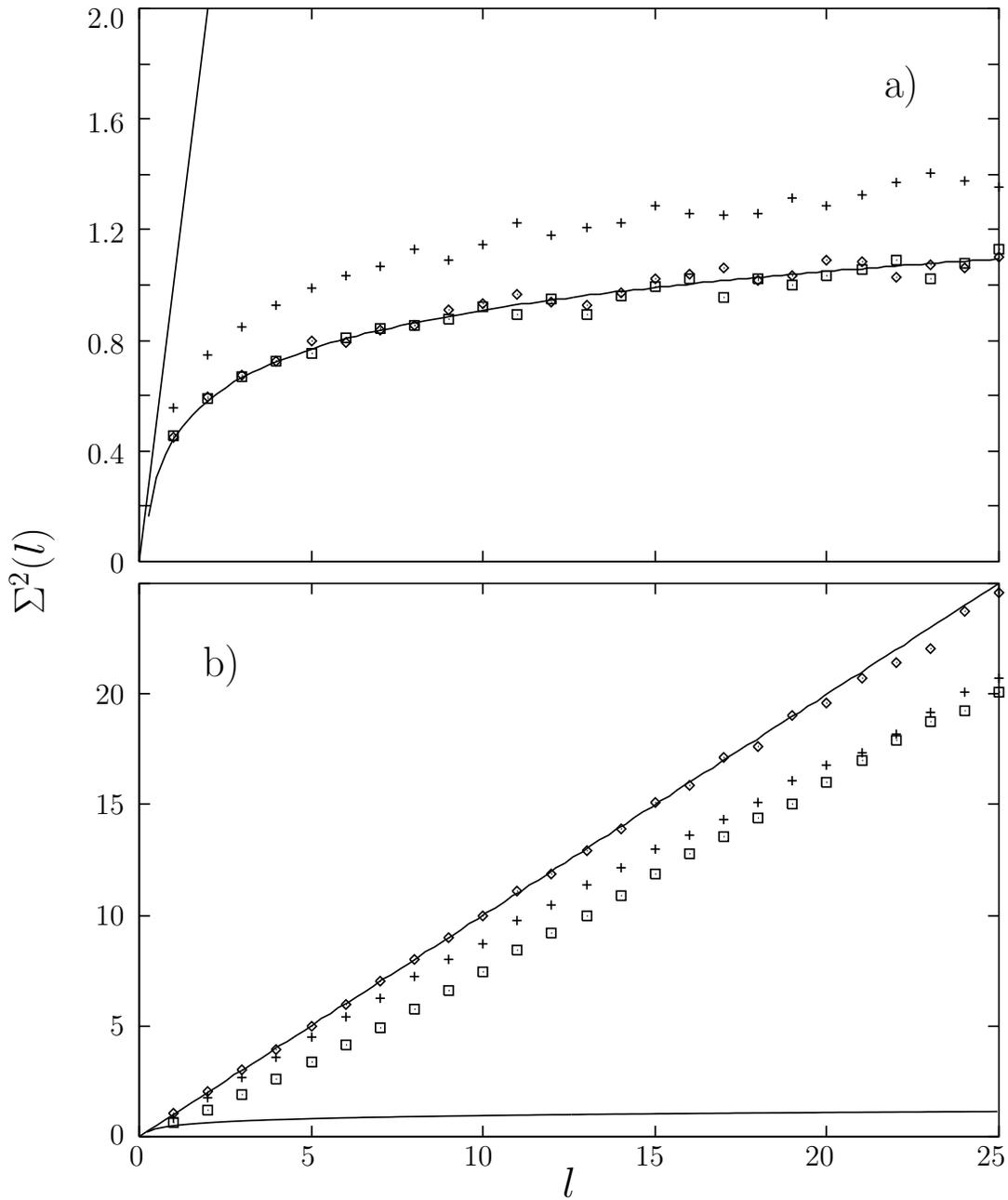}%
\end{picture}%
\setlength{\unitlength}{0.00058300in}%
\begingroup\makeatletter\ifx\SetFigFont\undefined
\def\x#1#2#3#4#5#6#7\relax{\def\x{#1#2#3#4#5#6}}%
\expandafter\x\fmtname xxxxxx\relax \def\y{splain}%
\ifx\x\y   
\gdef\SetFigFont#1#2#3{%
  \ifnum #1<17\tiny\else \ifnum #1<20\small\else
  \ifnum #1<24\normalsize\else \ifnum #1<29\large\else
  \ifnum #1<34\Large\else \ifnum #1<41\LARGE\else
     \huge\fi\fi\fi\fi\fi\fi
  \csname #3\endcsname}%
\else
\gdef\SetFigFont#1#2#3{\begingroup
  \count@#1\relax \ifnum 25<\count@\count@25\fi
  \def\x{\endgroup\@setsize\SetFigFont{#2pt}}%
  \expandafter\x
    \csname \romannumeral\the\count@ pt\expandafter\endcsname
    \csname @\romannumeral\the\count@ pt\endcsname
  \csname #3\endcsname}%
\fi
\fi\endgroup
\begin{picture}(9633,11574)(376,-11554)
\put(1456,-9916){\makebox(0,0)[rb]{\smash{\SetFigFont{12}{14.4}{rm}5}}}
\put(1456,-8851){\makebox(0,0)[rb]{\smash{\SetFigFont{12}{14.4}{rm}10}}}
\put(1456,-7771){\makebox(0,0)[rb]{\smash{\SetFigFont{12}{14.4}{rm}15}}}
\put(1456,-6706){\makebox(0,0)[rb]{\smash{\SetFigFont{12}{14.4}{rm}20}}}
\put(1456,-5431){\makebox(0,0)[rb]{\smash{\SetFigFont{12}{14.4}{rm}0}}}
\put(1456,-4366){\makebox(0,0)[rb]{\smash{\SetFigFont{12}{14.4}{rm}0.4}}}
\put(1456,-3301){\makebox(0,0)[rb]{\smash{\SetFigFont{12}{14.4}{rm}0.8}}}
\put(1456,-2221){\makebox(0,0)[rb]{\smash{\SetFigFont{12}{14.4}{rm}1.2}}}
\put(1456,-1156){\makebox(0,0)[rb]{\smash{\SetFigFont{12}{14.4}{rm}1.6}}}
\put(1576,-11191){\makebox(0,0)[b]{\smash{\SetFigFont{12}{14.4}{rm}0}}}
\put(3241,-11191){\makebox(0,0)[b]{\smash{\SetFigFont{12}{14.4}{rm}5}}}
\put(4891,-11191){\makebox(0,0)[b]{\smash{\SetFigFont{12}{14.4}{rm}10}}}
\put(6556,-11191){\makebox(0,0)[b]{\smash{\SetFigFont{12}{14.4}{rm}15}}}
\put(8206,-11191){\makebox(0,0)[b]{\smash{\SetFigFont{12}{14.4}{rm}20}}}
\put(9871,-11191){\makebox(0,0)[b]{\smash{\SetFigFont{12}{14.4}{rm}25}}}
\put(5776,-11491){\makebox(0,0)[b]{\smash{\SetFigFont{17}{20.4}{rm}$l$}}}
\put(9001,-811){\makebox(0,0)[b]{\smash{\SetFigFont{17}{20.4}{rm}a)}}}
\put(2401,-6436){\makebox(0,0)[b]{\smash{\SetFigFont{17}{20.4}{rm}b)}}}
\put(1456,-136){\makebox(0,0)[rb]{\smash{\SetFigFont{12}{14.4}{rm}2.0}}}
\put(1456,-10981){\makebox(0,0)[rb]{\smash{\SetFigFont{12}{14.4}{rm}0}}}
\put(676,-5836){\makebox(0,0)[lb]{\smash{
\SetFigFont{17}{20.4}{rm}$\Sigma^2(l)$
}}}
\end{picture}
\caption{Number variance $\Sigma^2(l)$ for three different values
of the overlapping parameter $\kappa= 0.1$ (diamonds), $\kappa= 1.0$ (crosses) and 
$\kappa= 10.$ (squares). a) POE and b) GOE. 
The solid lines show the theoretical curves: $\Sigma_{\rm G}^2$ (a) and 
$\Sigma_{\rm P}^2$ (b).}
\label{f_GP_sig2}
\end{figure}

\section{Correlations connected with the widths \label{s_CG}}

In \cite{Lom93} the number variance is generalized to an {\it intensity variance} by weighting
each level with the intensity of its line as it appears in the cross section. Here we do the
weighting with the widths of the resonances (in the one channel case both ways are identical).
We denote this quantity by $\Sigma_g^2(l)$. It gives the variance of the summed 
width in a given interval of length $l$.
Neglecting possible  correlations with the widths we relate $\Sigma_g^2(l)$
to the simple number variance.  Then we compare the numerical results with this 
formula. Occurring discrepancies indicate the existence of exactly those correlations
which had been neglected in the beginning.

Following \cite{Lom93}, we define the width--weighted stick spectrum
\be
\rho_g(x) = \sum y_i \delta(x-e_i),\qquad
\int \rho_g(x)\,{\rm d}x = N' 
\label{gl_rho_g}\ee
with the normalised widths $y_i = \Gamma_i/\langle\Gamma\rangle$.
For $\kappa >1$ the $K$ broad resonances are again skipped.
The summed width in an interval $\Delta$ of length $l$ is
\be
B(l) = \int_\Delta \rho_g(x)\;\d x 
\ee
and consequently the variance of the summed width is 
\be
\Sigma_g^2(l) = \langle B^2(l)\rangle - \langle  B(l)\rangle^2 .
\label{sigG1}
\ee
Now, we relate this quantity to the number variance $\Sigma^2(l)$ in
the following way:

\parbox{14.0cm}{
\begin{eqnarray*}
B^2(l) &=& \sum_{ij} y_i y_j
\int_{\Delta^2} \delta(x-e_i)\delta(y-e_j)\,\d x\d y\\
&=& \sum_i y_i^2 \left\{ \begin{array}{rl}
1 \quad &  E_i \in\Delta \\
0 \quad & \mbox{otherwise}
\end{array} \right.  + \sum_{i\ne j} y_i y_j
\int_{\Delta^2} \delta(x-e_i)\delta(y-e_j)\,\d x\d y .
\end{eqnarray*}
} \hfill \parbox{1cm}{\be \ee}
Assuming no  position--width correlations, one may perform the averages
in (\ref{sigG1}) separately :
\be
\langle B(l)\rangle = \langle y\rangle \; \langle n(l)\rangle
\label{gl_B}
\ee
and furthermore assuming no width--width correlations one obtains
\be
 \langle B^2(l)\rangle = \langle y^2\rangle\, \langle n(l)\rangle
  + \langle y\rangle^2
\sum_{i\ne j} \langle \int_{\Delta^2}
 \delta(x-e_i)\delta(y-e_j)\,\d x\d y\rangle.
\label{gl_Bsl}
\ee
It is $\langle n(l)\rangle = l$ because the mean level distance is one.
The corresponding expression without width--weighting is
\be
 \langle n^2(l)\rangle = \langle n(l)\rangle +
\sum_{i\ne j} \langle \int_{\Delta^2}
 \delta(x-e_i)\delta(y-e_j)\,\d x\d y\rangle.
\label{gl_Nsl}
\ee
Comparing (\ref{gl_Bsl}) and (\ref{gl_Nsl}) one arrives at
\be
\langle B^2(l)\rangle = \langle y^2\rangle \,l + \langle y\rangle^2
\left( \langle n^2(l)\rangle - l\right) 
\label{eq:B2}
\ee
and finally using (\ref{gl_S2l}) and (\ref{gl_B})
\be
\Sigma_g^2(l) = (\Delta y^2) l + \Sigma^2(l).
\label{gl28}
\ee
Here $(\Delta y^2)$ is the variance of the width distribution.
In the case that the normalised widths $y_i$ are $\chi_K^2$ distributed,
$(\Delta y^2) = 2/K$.

From (\ref{gl28}) we see, that we may clearly separate properties stemming 
from the correlations in the energy spectrum as well as from the width
distribution on one hand and correlations between energies and widths 
or between widths at different energies on the other hand. To achieve 
this we proceed \cite{Lom93} to create from the true
width--weighted spectrum (\ref{gl_rho_g}) a synthetic one
\[ \tilde\rho_g = \sum y_{\pi(i)} \delta(x-\eps_i) \]
where we apply a random permutation $\pi$ to the indices of the widths.
As (\ref{gl28}) only uses the independence of widths and levels the 
synthetic spectrum should obey this equation, while the true one $\rho_g$
may not if correlations exist.

Fig. 6 shows the theoretical curves 
$\Sigma^2_{\rm P}$ (\ref{gl_S2P}) and $\Sigma^2_{\rm G}$ (\ref{gl_S2G})
as well as $\Sigma^2$ for the numerically obtained energy spectrum 
for 3 channels and $\kappa=100$.
We furthermore show $\Sigma^2_g-(\Delta y^2)l$ for both the true and the synthetic
(decorrelated) width--weighted stick spectra under the same conditions $K=3$ and $\kappa=100$.
The result for the synthetic spectrum agrees very well with that for the energy 
spectrum as it should be according to (\ref{gl28}).
The $\Sigma^2_g$ obtained for the true spectrum on the other hand differs greatly. 
This is a clear indication that the spectrum in this case presents 
strong correlations, which must be contrasted to the chaotic case where a
similar analysis indicates the absence of correlations. Note however that the
agreement of $\Sigma^2_g-(\Delta y^2)l$ with $\Sigma_{\rm G}^2$ for $l<5$
is only accidental, what can be concluded from the results for
different number of channels in Fig.\,7.

The nature of these correlations is such that the spectrum becomes more rigid due
to the width--weighting (disregarding the term $(\Delta y^2) l$). This may be understood 
from (\ref{gl_gdis1}) meaning that the width of a trapped resonance is
more likely to be large if the distance between its neighboring levels is large, too
$\Gamma_\nu \sim |\eps_\nu - \eps_{\nu-1}|^2$. \\

\begin{figure}
\begin{picture}(0,0)%
\epsfbox{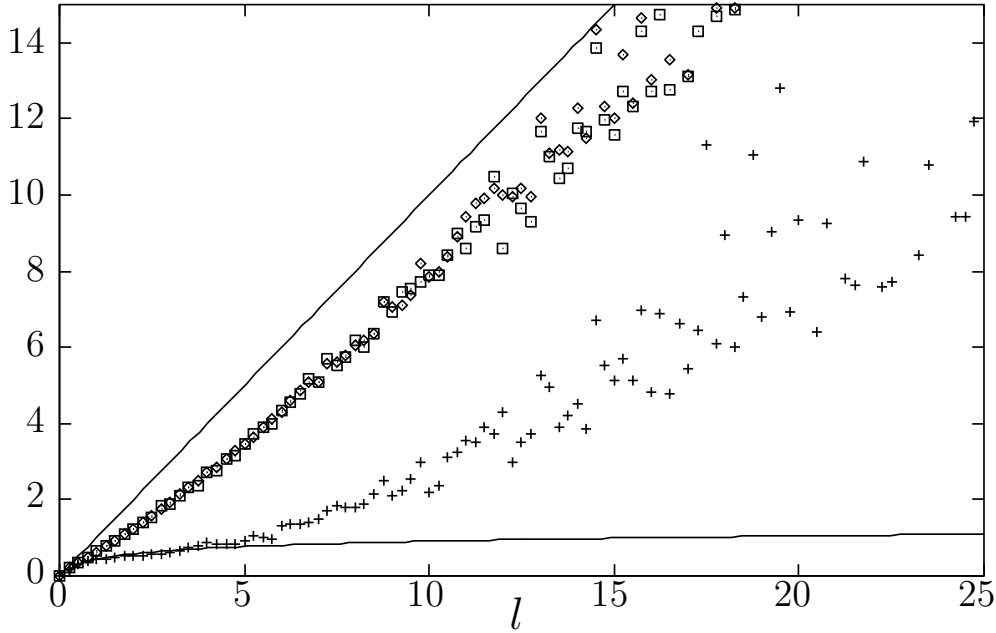}%
\end{picture}%
\setlength{\unitlength}{0.00058300in}%
\begingroup\makeatletter\ifx\SetFigFont\undefined
\def\x#1#2#3#4#5#6#7\relax{\def\x{#1#2#3#4#5#6}}%
\expandafter\x\fmtname xxxxxx\relax \def\y{splain}%
\ifx\x\y   
\gdef\SetFigFont#1#2#3{%
  \ifnum #1<17\tiny\else \ifnum #1<20\small\else
  \ifnum #1<24\normalsize\else \ifnum #1<29\large\else
  \ifnum #1<34\Large\else \ifnum #1<41\LARGE\else
     \huge\fi\fi\fi\fi\fi\fi
  \csname #3\endcsname}%
\else
\gdef\SetFigFont#1#2#3{\begingroup
  \count@#1\relax \ifnum 25<\count@\count@25\fi
  \def\x{\endgroup\@setsize\SetFigFont{#2pt}}%
  \expandafter\x
    \csname \romannumeral\the\count@ pt\expandafter\endcsname
    \csname @\romannumeral\the\count@ pt\endcsname
  \csname #3\endcsname}%
\fi
\fi\endgroup
\begin{picture}(8830,5710)(1179,-5779)
\put(1456,-511){\makebox(0,0)[rb]{\smash{\SetFigFont{14}{16.8}{rm}14}}}
\put(5716,-5716){\makebox(0,0)[b]{\smash{\SetFigFont{17}{20.4}{rm}$l$}}}
\put(1576,-5506){\makebox(0,0)[b]{\smash{\SetFigFont{14}{16.8}{rm}0}}}
\put(3241,-5506){\makebox(0,0)[b]{\smash{\SetFigFont{14}{16.8}{rm}5}}}
\put(4891,-5506){\makebox(0,0)[b]{\smash{\SetFigFont{14}{16.8}{rm}10}}}
\put(6556,-5506){\makebox(0,0)[b]{\smash{\SetFigFont{14}{16.8}{rm}15}}}
\put(8206,-5506){\makebox(0,0)[b]{\smash{\SetFigFont{14}{16.8}{rm}20}}}
\put(9871,-5506){\makebox(0,0)[b]{\smash{\SetFigFont{14}{16.8}{rm}25}}}
\put(1456,-5296){\makebox(0,0)[rb]{\smash{\SetFigFont{14}{16.8}{rm}0}}}
\put(1456,-4606){\makebox(0,0)[rb]{\smash{\SetFigFont{14}{16.8}{rm}2}}}
\put(1456,-3931){\makebox(0,0)[rb]{\smash{\SetFigFont{14}{16.8}{rm}4}}}
\put(1456,-3241){\makebox(0,0)[rb]{\smash{\SetFigFont{14}{16.8}{rm}6}}}
\put(1456,-2566){\makebox(0,0)[rb]{\smash{\SetFigFont{14}{16.8}{rm}8}}}
\put(1456,-1876){\makebox(0,0)[rb]{\smash{\SetFigFont{14}{16.8}{rm}10}}}
\put(1456,-1186){\makebox(0,0)[rb]{\smash{\SetFigFont{14}{16.8}{rm}12}}}
\end{picture}
\caption{Variance $\Sigma_g^2(l)$ of the summed width subtracted by
$(\Delta y^2) l$ for the POE case. The crosses denote this quantity 
for the original spectrum
which contains all correlations. The squares correspond to the shuffled spectrum  and
the diamonds denote the pure number variance $\Sigma^2$. The solid lines show the 
theoretical curves $\Sigma^2_{\rm G}$ and $\Sigma^2_{\rm P}$.}
\label{f_sigg2}
\end{figure}

Furthermore we expect that for increasing number $K$ of channels the effect of 
the position -- width and the width -- width correlations vanishes, because
the variance of the width distribution $2/K$ becomes very small. In this case 
the sum of the normalised widths becomes more and more equal to the number of levels in 
the interval considered. This is verified in Fig.\,7.

\begin{figure}
\begin{picture}(0,0)%
\epsfbox{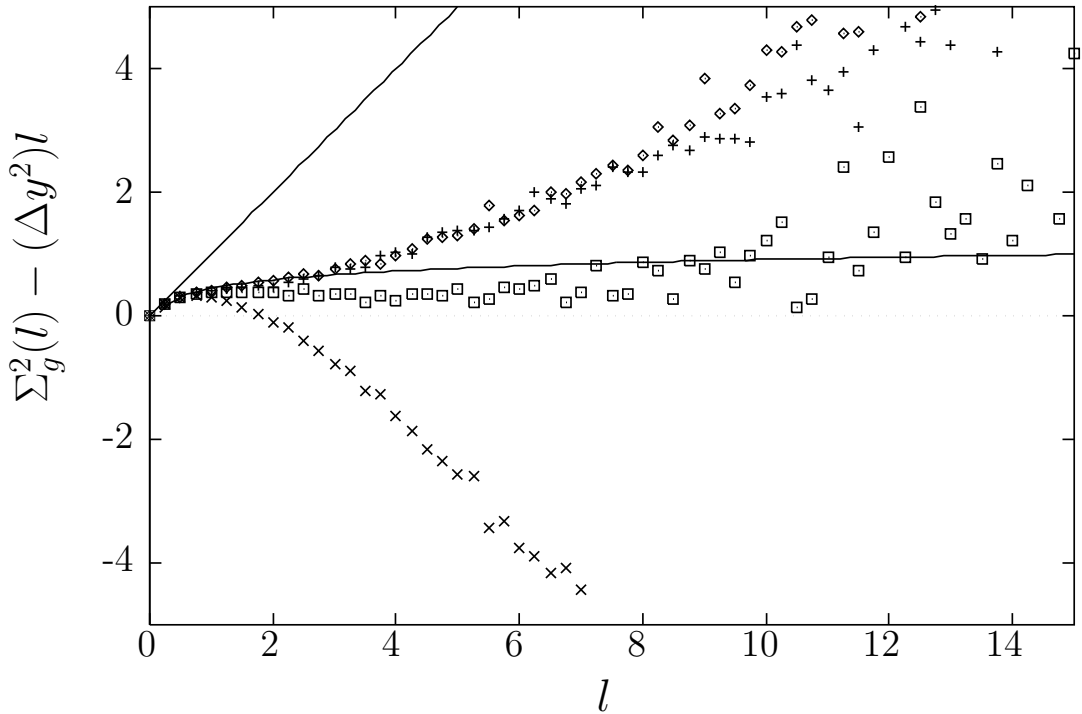}%
\end{picture}%
\setlength{\unitlength}{0.00058300in}%
\begingroup\makeatletter\ifx\SetFigFont\undefined
\def\x#1#2#3#4#5#6#7\relax{\def\x{#1#2#3#4#5#6}}%
\expandafter\x\fmtname xxxxxx\relax \def\y{splain}%
\ifx\x\y   
\gdef\SetFigFont#1#2#3{%
  \ifnum #1<17\tiny\else \ifnum #1<20\small\else
  \ifnum #1<24\normalsize\else \ifnum #1<29\large\else
  \ifnum #1<34\Large\else \ifnum #1<41\LARGE\else
     \huge\fi\fi\fi\fi\fi\fi
  \csname #3\endcsname}%
\else
\gdef\SetFigFont#1#2#3{\begingroup
  \count@#1\relax \ifnum 25<\count@\count@25\fi
  \def\x{\endgroup\@setsize\SetFigFont{#2pt}}%
  \expandafter\x
    \csname \romannumeral\the\count@ pt\expandafter\endcsname
    \csname @\romannumeral\the\count@ pt\endcsname
  \csname #3\endcsname}%
\fi
\fi\endgroup
\begin{picture}(9562,6415)(376,-6274)
\put(676,-3436){\makebox(0,0)[lb]{\smash{
\SetFigFont{17}{20.4}{rm}$\Sigma^2_g(l) - (\Delta y^2) l$
}}}
\put(1456,-511){\makebox(0,0)[rb]{\smash{\SetFigFont{14}{16.8}{rm}4}}}
\put(1456,-1621){\makebox(0,0)[rb]{\smash{\SetFigFont{14}{16.8}{rm}2}}}
\put(1456,-2731){\makebox(0,0)[rb]{\smash{\SetFigFont{14}{16.8}{rm}0}}}
\put(1456,-3841){\makebox(0,0)[rb]{\smash{\SetFigFont{14}{16.8}{rm}-2}}}
\put(1456,-4951){\makebox(0,0)[rb]{\smash{\SetFigFont{14}{16.8}{rm}-4}}}
\put(1576,-5716){\makebox(0,0)[b]{\smash{\SetFigFont{14}{16.8}{rm}0}}}
\put(2686,-5716){\makebox(0,0)[b]{\smash{\SetFigFont{14}{16.8}{rm}2}}}
\put(3781,-5716){\makebox(0,0)[b]{\smash{\SetFigFont{14}{16.8}{rm}4}}}
\put(4891,-5716){\makebox(0,0)[b]{\smash{\SetFigFont{14}{16.8}{rm}6}}}
\put(6001,-5716){\makebox(0,0)[b]{\smash{\SetFigFont{14}{16.8}{rm}8}}}
\put(7111,-5716){\makebox(0,0)[b]{\smash{\SetFigFont{14}{16.8}{rm}10}}}
\put(8206,-5716){\makebox(0,0)[b]{\smash{\SetFigFont{14}{16.8}{rm}12}}}
\put(9316,-5716){\makebox(0,0)[b]{\smash{\SetFigFont{14}{16.8}{rm}14}}}
\put(5701,-6211){\makebox(0,0)[b]{\smash{\SetFigFont{17}{20.4}{rm}$l$}}}
\end{picture}
\caption{Variance $\Sigma_g^2(l)$ 
of the summed width subtracted by $(\Delta y^2) l$
but only for the original spectrum, containing all correlations. The diamonds 
denote the case $K=6$, the upright crosses $K=3$, the squares $K=2$ and the crosses 
$K=1$. The solid lines show again $\Sigma^2_{\rm G}$ and $\Sigma^2_{\rm P}$.}
\label{f_Ksigg2}
\end{figure}

\section{Summary}

We considered the distortion of a regular system by its coupling $\kappa$ 
to the continuum of decay channels. We received results for 

\begin{itemize}
\item{\ldots the width  distribution, as it is altered with increasing
$\kappa$:  In the weak coupling limit for both the GOE and the POE, the
widths are $\chi_K^2$ distributed when normalised to unit mean. Then with
increasing $\kappa$ their distributions become broader. For the GOE, the
widths return to their original distribution in the strong coupling limit.
For the POE, the width distribution becomes approximately a $\chi_q^2$
distribution again, but  in contrast to the GOE with much larger variance
as before (cf. Fig.\,2 and Fig.\,3).}

\item{\ldots the correlations
between the positions alone: For this analysis we used the number variance 
$\Sigma^2$. For the GOE we found correlations as they are typical for the level
statistics  of a closed system in both the weak and the strong coupling region.
Only in the critical region the correlations become weaker. The
POE starts  without correlations. Then they increase steadily with the
coupling parameter $\kappa$ (cf. Fig.\,5).}

\item{\ldots  the correlations connected with the widths: Here we used the
generalized measure $\Sigma_g^2$ \cite{Lom93}, calculated from a
width--weighted stick spectra.
There appear two additional types of correlations, namely such
between the position and the width of each resonance and such between the
widths of different resonances. In the GOE case they appear only in the
critical region, whereas in the POE case they increase steadily (cf. Fig.\,6 and 7).}

\end{itemize}

Furthermore we derived an
analytic expression for the width distribution of the GOE and POE
in the one channel case at strong coupling (cf. Fig.\,4).
It relates the width of a trapped state to the distance of the two neighbored levels and
explains by this the different width distribution of GOE and POE.\\

The results of our investigations show the special role of the GOE.
Its properties survive the distortion of the system by coupling it to the continuum:
at large coupling the correlations and the width distribution
are the same as at low coupling. In contrast to the GOE, the 
properties of the POE are not restored at strong coupling strength.\\

Realistic systems are often in the critical region where correlations in the spectrum
are induced by the coupling to the continuum. 
Under these conditions the S--matrix poles are difficult to find. Nevertheless they
determine the statistical properties of the cross section.
We will investigate this problem in a forthcoming paper for both the GOE and the POE.
\vspace{.8cm}

{\small {\bf Acknowledgment:} Valuable discussions with G. Soff and V. V. Sokolov are
greatfully acknowledged. The investigations are supported by DFG (Ro 922/6) and by DAAD.}

\vspace{1cm}
\begin{appendix}

\section{Resonance width in the two level approximation \label{as_W}}

Writing (\ref{gl_polesc}) for two neighboring levels, we get
\be
\frac{v_1^2}{s+\frac{\Delta}{2}} + \frac{v_2^2}{s-\frac{\Delta}{2}} = 0, \qquad
\frac{v_1^2}{(s+\frac{\Delta}{2})^2} + \frac{v_2^2}{(s-\frac{\Delta}{2})^2} = 
\frac{4}{\eta\Gamma}
\label{gl_ap1}
\ee
where $\Delta= \eps_2 -\eps_1$ is the distance between the levels and $E$ is 
substituted by $E= \frac{1}{2}(\eps_1+ \eps_2) +s$. Due to the first equation
\be
s= - \frac{\Delta}{2} \frac{v_2^2 -v_1^2}{v_1^2 +v_2^2} = -\frac{\Delta}{2} \tau.
\ee
Inserting this result into the second equation of (\ref{gl_ap1}) yields
\be
\frac{\Delta^2}{\eta\Gamma} = \left( \frac{v_1^2}{(1-\tau)^2} +
\frac{v_2^2}{(1+\tau)^2} \right) = \frac{v_1^2+v_2^2}{1-\tau^2}.
\ee
Finally measuring $\Delta$ in units of the mean level spacing $\Delta= d_0 s$ in the center 
of the spectrum, and $\Gamma$ in units of the mean width
$ \Gamma= \langle \Gamma_\nu\rangle y$ according to (\ref{gl_gamf}), 
one arrives at (\ref{gl_gdis1})
\be
y= \frac{\pi^2 s^2}{4N(v_1^2+v_2^2)} ( 1-\tau^2).
\ee

\section{Width distribution \label{as_WD}}

The width distribution according to (\ref{gl_gdis1}) is
\be
p(y) = \frac{N}{2\pi} \int \d v_1\,\d v_2\,{\rm e}^{\T -\frac{N}{2}(v_1^2+v_2^2)}\;
\int \d s\, P(s)\, \delta\left[ y- \frac{\pi^2 s^2}{4N(v_1^2+v_2^2)} \left( 1-\tau^2\right)
\right] .
\ee
Transforming $v_1$ and $v_2$ into spherical coordinates $r/{\T\sqrt{N}} \cos\phi$ and
$r/{\T\sqrt{N}} \sin\phi$ so that $\tau= \sin^2\phi -\cos^2\phi= -\cos 2\phi$ we get
\be
p(y) = \frac{1}{2\pi} \int r\d r\,\d \phi\,{\rm e}^{\T -\frac{r^2}{2}}\;
\int \d s\, P(s)\,\delta\left[ y- \frac{\pi^2 s^2 \sin^2 2\phi}{4r^2}\right].
\ee
It is enough to integrate $\phi$ from $0$ to $\pi/4$ because of the eight--fold symmetry 
of the integrand. Applying in addition the transformation $x= \sin^2 2\phi$
\be
p(y)= \frac{1}{\pi} \int r\d r \,{\rm e}^{\T -\frac{r^2}{2}}\;
\int_0^1 \frac{\d x}{\sqrt{x(1-x)}}\; 
\int \d s\, P(s)\, \delta\left[ y- \frac{\pi^2 s^2 x}{4r^2}\right].
\label{gl_gdisa}\ee
In order to go ahead we consider the two cases GOE and POE separately in the 
following subsections.

\subsection{GOE case}

Here the nearest neighbor distribution reads
\be
P_{\rm G}(s) = \frac{\pi}{2}s \, {\rm e}^{-\T \frac{\pi}{4}s^2}.
\ee
In order to resolve the $\delta$ function, we substitute $r$ as a function of 
$y'= \frac{\pi^2 s^2 x}{4r^2}$ and use $\int\delta(y-y') f(y') \d y = f(y)$. This 
leads to
\be
p_{\rm G}(y) = \frac{-\pi^2}{16 y^2} \int_0^1 \d x\, \sqrt{\frac{x}{1-x}}\;
\int_0^\infty \d s \,s^3 {\rm e}^{-\T \alpha s^2},\qquad 
\alpha= \frac{\pi}{4}(1+\frac{\pi x}{2y}).
\ee
The integral over the level spacing $s$ gives $-1/(2\alpha^2)$ and it remains a last
integration, namely
\be
p_{\rm G}(y) = \frac{2}{\pi^2} \int_0^1 \d x\, \sqrt{\frac{x}{1-x}}\,
\left(\frac{2y}{\pi} +x\right)^{-2} .
\ee
This may be solved by substituting $t= \sqrt{\frac{x}{1-x}}$ 
\be
p_{\rm G}(y) = \frac{4}{\pi^2 (1+b)^2} \int_0^\infty \d t \,
\frac{t^2}{(t^2+ \frac{b}{b+1})^2},\qquad b= \frac{2y}{\pi}
\ee
and integrating by parts. Then
\be
p_{\rm G}(y) = \frac{1}{\pi(1+b)^2} \sqrt{\frac{1+b}{b}} = \frac{1}{\sqrt{2\pi y}}
\left(1+\frac{2y}{\pi}\right)^{-\frac{3}{2}} .
\ee

\subsection{POE case}

Here the nearest neighbor distribution is
\be
P_{\rm P}(s) = {\rm e}^{-s}.
\ee
In contrast to the GOE case we first substitute $s$ instead of $r$. It follows that
\begin{eqnarray*}
p_{\rm P}(s) &=& \frac{1}{\pi^2 \sqrt{y}} \int r^2 \d r \,{\rm e}^{\T -\frac{r^2}{2}}\;
\int_0^1 \frac{\d x}{x\sqrt{1-x}}\;
{\rm e}^{-\T \frac{2r}{\pi}\sqrt{\frac{y}{x}}} \\
&=& \frac{2}{\pi^2 \sqrt{y}} \int\d r\, r^2\,{\rm e}^{\T -\frac{r^2}{2}}\;
\int_1^\infty \frac{ \d z\, {\rm e}^{-\alpha z}}{\sqrt{z^2-1}} ,\qquad 
\alpha= \frac{2r}{\pi}\sqrt{y} .
\end{eqnarray*}
The last integral represents the modified Bessel function 
\cite{ASt64}
\be
p_{\rm P}(s) = \frac{2}{\pi^2 \sqrt{y}} \int\d r\, r^2\,{\rm e}^{\T -\frac{r^2}{2}}\;
K_0({\T\frac{2\sqrt{y}}{\pi}} r)
\ee
This integral can be found in \cite{GRy80}
\be
p_{\rm P}(s) = \frac{1}{4y} {\rm e}^{\frac{y}{\pi^2}} W_{-1,0}(\T\frac{2y}{\pi^2}) .
\ee

\end{appendix}


\newpage
\begin{thebibliography}{10}

\bibitem{Fes58}
H.~Feshbach.
\newblock {\em Ann. Phys. (N.Y.)}, {\bf 5}, 357\, (1958).

\bibitem{Fes62}
H.~Feshbach.
\newblock {\em Ann. Phys. (N.Y.)}, {\bf 19}, 287\, (1962).

\bibitem{Mah69}
C.~Mahaux and H.~A. Weidenm\"uller.
\newblock {\em Shell--Model Approach to Nuclear Reactions}.
\newblock (North--Holland Publishing Company, 1969).

\bibitem{Kle85}
P.~Kleinw\"achter and I.~Rotter.
\newblock {\em Phys. Rev. C}, {\bf 32}, 1742\, (1985).

\bibitem{Sok89}
V.~V. Sokolov and V.~G. Zelevinsky.
\newblock {\em Nucl. Phys. A}, {\bf 504}, 562\, (1989).

\bibitem{Haa92}
F.~Haake, F.~Izrailev, N.~Lehmann, D.~Saher, and H.-J. Sommers.
\newblock {\em Z. Phys. B}, {\bf 88}, 359\, (1992).

\bibitem{Bro81}
T.~A. Brody, J.~Flores, J.~B. French, P.~A. Mello, A.~Pandey, 
and S.~S.~M. Wong.
\newblock {\em Rev. Mod. Phys.}, {\bf 53}(3), 385\, (1981).

\bibitem{Por65}
C.~E. Porter, editor.
\newblock {\em Statistical theories of spectra: Fluctuations}.
\newblock (Academic Press, New York, 1965).

\bibitem{Mue87}
A.~M\"uller and H.~L. Harney.
\newblock {\em Phys. Rev. C}, {\bf 35}(4), 1228\, (1987).

\bibitem{Dit91b}
F.-M. Dittes, I.~Rotter, and T.~H. Seligman.
\newblock {\em Phys. Lett. A}, {\bf 158}, 14\, (1991).

\bibitem{Seb96}
S.~Albeverio, F.~Haake, P.~Kurasov, M.~Ku\'{s}, and P.~\v{S}eba.
\newblock {\em J. Math. Phys.}, 1997 (in press).

\bibitem{Sto90}
H.-J. St\"ockmann and J.~Stein.
\newblock {\em Phys. Rev. Lett.}, {\bf 64}(19), 2215\, (1990).

\bibitem{Sto91}
F.~Haake, G.~Lenz, P.~\v{S}eba, J.~Stein, H.-J. St\"ockmann, and
K.~\.Zyczkowski.
\newblock {\em Phys. Rev. A}, {\bf 44}(10), R6161\, (1991).

\bibitem{Mar93}
C.~M. Marcus, R.~M. Westervelt, P.~F. Hopkins, and A.~C. Gossard.
\newblock {\em CHAOS}, {\bf 3}(4), 643\, (1993).

\bibitem{Jun97}
C.~Jung and T.~H. Seligman.
\newblock {\em Phys. Reps.}, 1997 (in press).

\bibitem{Alh86}
Y.~Alhassid and R.~D. Levine.
\newblock {\em Phys. Rev. Lett.}, {\bf 57}(23), 2879\, (1986).

\bibitem{Alh89}
Y.~Alhassid and M.~Feingold.
\newblock {\em Phys. Rev. A}, {\bf 39}(1), 374\, (1989).

\bibitem{Dit92}
F.-M. Dittes, H.~L. Harney, and A. M\"uller.
\newblock {\em Phys. Rev. A}, {\bf 45}(2), 701\, (1992).

\bibitem{Per96}
E.~Persson, T.~Gorin, and I.~Rotter.
\newblock {\em Phys. Rev. E}, {\bf 54}(4), 3339\, (1996).

\bibitem{Fyo96}
Y.~V. Fyodorov and H.-J. Sommers.
\newblock {\em JETP Letters}, {\bf 63}, 970\, (1996).

\bibitem{Boh89}
O.~Bohigas.
\newblock Chaos and quantum physics.
\newblock In M.~J. Giannoni, editor, {\em Proceedings of the Les 
Houches Summer School. Session LII}, page~91. (North Holland, 
Amsterdam, 1991).

\bibitem{Rot91}
I.~Rotter.
\newblock {\em Rep. Prog. Phys.}, {\bf 54}, 635\, (1991).

\bibitem{Sok92}
V.~V. Sokolov and V.~G. Zelevinsky.
\newblock {\em Ann. Phys.}, {\bf 216}(2), 323\, (1992).

\bibitem{Lor89}
F.~Lorenz.
\newblock {\em Lineare Algebra II}.
\newblock (BI--Wissenschaftsverlag, 1989).

\bibitem{Ver85}
J.~J.~M. Verbaarschot, H.~A. Weidenm\"uller, and M.~R. Zirnbauer.
\newblock {\em Phys. Reps.}, {\bf 129}(6), 367\, (1985).

\bibitem{GRy80}
I.~S. Gradshteyn and I.~M. Ryzhik.
\newblock {\em Table of Integrals, Series and Products}.
\newblock (Academic Press, San Diego, 1980).

\bibitem{ASt64}
M.~Abramowitz and I.~A. Stegun, editors.
\newblock {\em Handbook of mathematical functions}.
\newblock (Dover Publications, INC., New York\, 1970).

\bibitem{Miz93}
S.~Mizutori and V.~G. Zelevinsky.
\newblock {\em Z. Phys. A}, {\bf 346}, 1\, (1993).

\bibitem{Mue95}
M.~M\"uller, F.-M.~Dittes, W.~Iskra, and I.~Rotter.
\newblock {\em Phys. Rev. E}, {\bf 52}(6), 5961\, (1995).

\bibitem{Lom93}
M.~Lombardi and T.~H. Seligman.
\newblock {\em Phys. Rev. A}, {\bf 47}(5), 3571\, (1993).

\end{thebibliography}
\end{document}